\documentclass[a4paper]{jpconf}
\usepackage{graphicx}
\newcommand{\nnbe}{\stackrel{{\scriptscriptstyle (} - {\scriptscriptstyle )}}{\nu}\!\!\!\! _{\rm e}}
\begin{document}
\title{Capturing Relic Neutrinos with $\beta$--decaying Nuclei}

\author{Alfredo G. Cocco$^1$, Gianpiero Mangano$^{1,2}$, Marcello Messina$^3$}
\address{$^1$ INFN, Sezione di Napoli, Monte S.Angelo I-80126, Naples, Italy}
\address{$^2$ LAPTH, Universit\'{e} de Savoie et CNRS, F-64941,
Annecy-le-vieux Cedex, France}
\address{$^3$ Laboratorium f\"ur Hochenergiephysik - Universit\"at Bern, CH-3012, Bern, Switzerland}
\ead{cocco@na.infn.it, mangano@na.infn.it,
marcello.messina@cern.ch}

\begin{abstract}
We summarize a novel approach which has been recently proposed for
direct detection of low energy neutrino backgrounds such as the
cosmological relic neutrinos, exploiting neutrino/antineutrino
capture on nuclei that spontaneously undergo $\beta$ decay.
\end{abstract}
\section{Introduction}
The electron {\bf N}eutrino {\bf C}apture from a nucleus $A$ which
spontaneously decays via {\bf B}eta decay to a daughter nucleus
$B$ (in the following NCB)
\begin{equation}
\nnbe + A \rightarrow B + e^\pm \label{main} \, ,
\end{equation}
shows the remarkable property that it has no energy threshold on
the value of the incoming neutrino energy. In the limit of
vanishing value for neutrino mass $m_\nu$ and energy the neutrino
contributes to (\ref{main}) uniquely via its lepton flavor quantum
number and in this case the electron in the final state has
exactly the $\beta$ decay endpoint energy $Q_\beta$. However, for
finite $m_\nu$ the electron kinetic energy is $Q_\beta+E_\nu \geq
Q_\beta + m_\nu$, while electrons emerging from the corresponding
$\beta$ decay has at most an energy $Q_\beta - m_\nu$, neglecting
nucleus recoil energy.

The idea of using NCB to measure the cosmological relic neutrino
background predicted in the framework of the Hot Big Bang model
was already suggested many years ago in \cite{weinberg}. The
original idea was that if relic neutrinos have a large chemical
potential $\mu$, the electron (positron) energy spectrum for
$\beta$ decays and NCB would get quite a typical signature in a
interval of order $\mu$ around the zero neutrino mass endpoint
$Q_\beta$. However, Big Bang Nucleosynthesis constrains relic
neutrino--antineutrino asymmetry to the small value $\mu/T_\nu
\leq 0.1$, see e.g. \cite{cuoco,hansen}. This implies that, unless
more exotic scenarios are considered, as a larger amount of
relativistic degrees of freedom in the Early Universe, the effect
of neutrino degeneracy in $\beta$ decays and NCB is too small to
be detected experimentally. However, for massive neutrinos a gap
around $Q_\beta$ is expected of the order of twice the neutrino
mass, which for $m_\nu \sim 1$ eV is several orders of magnitude
larger than the corresponding effect due to neutrino-antineutrino
asymmetry. At least in principle, this allows to distinguish
between $\beta$ decay and NCB interaction. In this paper we
briefly summarize the results of \cite{cocco} where it is argued
that if $m_\nu$ is in the eV range, future NCB experiments could
represent an almost unique way to detect cosmological neutrinos.
\section{The Neutrino Capture Rate}
Assuming an isotropic neutrino flux corresponding to a
distribution function in phase space $f(p_\nu)$, the NCB
integrated rate can be expressed as an integral over the electron
(positron) energy
\begin{equation}
\lambda_\nu = \int \sigma_{\scriptscriptstyle {\rm NCB}} v_\nu
\,f(p_\nu) \, \frac{d^3 p_\nu}{(2 \pi)^3} = \frac{G_\beta^2}{2
\pi^3} \int_{W_o+2 m_\nu}^\infty p_e E_e F(Z,E_e) C(E_e,p_\nu)_\nu
E_\nu p_\nu \,f(p_\nu) \, d E_e \, , \label{sigma1bis}
\end{equation}
where $F(Z,E_e)$ is the Fermi function, with $E_e$ and $p_e$ the
energy and momentum of the outgoing electron and $W_o$ the
corresponding $\beta$ decay endpoint. The rate contains the
nuclear shape factor $C(E_e,p_\nu)_\nu$, an angular momentum
weighted average of nuclear state transition amplitudes, which
depends upon the nuclear properties of the parent and daughter
nuclei and represents the main source of uncertainty in
$\sigma_{\scriptscriptstyle {\rm NCB}} v_\nu$, the product of the
cross section times neutrino velocity.

On the other hand, NCB rate is strongly related to the
corresponding $\beta$ decay process rate
\begin{equation}
\lambda_\beta =   \frac{G_\beta^2}{2 \pi^3} \int_{m_e}^{W_o} p_e
E_e F(Z,E_e) C(E_e,p_\nu)_\beta E_\nu p_\nu \, d E_e \, ,
\label{ratedecay}
\end{equation}
where a simple relation between the two shape factors holds
\begin{equation}
C(E_e,p_\nu)_\nu = C(E_e,-p_\nu)_\beta \, , \label{cpos}
\end{equation} though both variables have different kinematical
domains in the two processes.

Therefore, the $\beta$ decay rate can be used to provide a
relation giving the mean shape factor, defined as
\begin{equation}
\overline{C}_\beta = \frac{1}{f} \int_{m_e}^{W_o} p_e E_e F(Z,E_e)
C(E_e,p_\nu)_\beta E_\nu p_\nu  dE_e \, ,
\end{equation}
in terms of observable quantities, $W_o$ and the product $f
t_{1/2}$
\begin{equation}
ft_{1/2} = \frac{2\pi^3 \ln2}{G_\beta^2 \ \overline{C}_\beta} \, ,
 \label{ft}
\end{equation}
Defining
\begin{equation}
{\cal A} = \int_{m_e}^{W_o}
\frac{C(E'_e,p'_\nu)_\beta}{C(E_e,p_\nu)_\nu} \frac{p'_e}{p_e}
\frac{E'_e}{E_e} \frac{F(E'_e, Z)}{F(E_e, Z)} E'_\nu p'_\nu dE'_e
\, , \label{aint}
\end{equation}
where a prime denotes all variables depending on $E'_e$ which
should be integrated over, the NCB cross section times neutrino
velocity can then be conveniently written as
\begin{equation}
\sigma_{\scriptscriptstyle {\rm NCB}} v_\nu= \frac{2\pi^2 \ln2
}{{\cal A}\cdot  t_{1/2}} \, , \label{sigma3}
\end{equation}

Notice that ${\cal A} $ contains the ratio of NCB and $\beta$
decay shape factors. As discussed in details in \cite{cocco}, in
several relevant cases (super-allowed transitions, unique k-th
forbidden transitions) the evaluation of ${\cal A}$ is
particularly simple so that Eq. (\ref{sigma3}) can be computed in
an exact way. Furthermore, in all cases where this is not
possible, systematic uncertainties affecting the nuclear matrix
element evaluation largely cancel in the shape factor ratio
appearing in ${\cal A}$, thus providing a reliable estimate of the
NCB cross sections. Results for both allowed and unique forbidden
decay cross sections having branching ratios greater than $5\%$,
namely 1272 $\beta^-$ decays and 799 $\beta^+$ decays can be found
in \cite{cocco}. Indeed, there are several nuclei spanning a wide
range in $Q_\beta$ for which interesting high values are reached
in the range $\sigma_{\scriptscriptstyle {\rm NCB}} (v_\nu/c) =
10^{-41}-10^{-43}$ cm$^2$. As an example, in the interesting case
of $^3$H one gets $\sigma_{\scriptscriptstyle {\rm NCB}} (v_\nu/c)
= 7.84 \times 10^{-45}$ cm$^2$, which for the standard homogeneous
flux of cosmological neutrinos corresponds to 7.5 events per year
of data taking for a mass of 100 g. In general, this estimate
represents a lower bound, as massive neutrino density is expected
to be locally larger because of gravitational clustering. This
effect in a Cold Dark Matter Halo is quite relevant for order eV
neutrino masses, see Table \ref{table:1}.

Of course, the finite energy resolution of any experimental
apparatus and the extremely low cross section make relic neutrino
detection via NCB a real challenge due to the large background
events produced by standard $\beta$ decay. In particular, the
ratio of the event rate $\lambda_\beta(\Delta)$ for the last
$\beta$ decay electron energy bin $W_o-\Delta<E_e<W_o$, compared
with the total NCB event rate is typically very large, since
$\Delta >> T_\nu$
\begin{equation}
\frac{\lambda_\beta(\Delta)}{\lambda_\nu} = \frac{2}{9 \zeta(3)}
\left( \frac{\Delta}{T_\nu} \right)^3 \left(1+ \frac{2
m_\nu}{\Delta} \right)^{3/2} >> 1 \, ,
\end{equation}
It is therefore a crucial issue to reach an energy resolution, the
electron energy bin dimension of the apparatus $\Delta$, smaller
than $ m_\nu$. For example, the expected background electron
events which are produced by $\beta$ decay, yet having an energy
which corresponds to the relic neutrino capture energy bin
centered at $E_e=W_o+2 m_\nu$ are smaller than a factor three with
respect to NCB processes if $\Delta= 0.2$ eV for $m_\nu = 0.7$ eV,
while a smaller neutrino mass of $0.3$ eV requires $\Delta=0.1$
eV. Presently, to obtain such an energy resolution seems very
demanding. Nevertheless, if a large neutrino mass will be found by
the ongoing $\beta$ decay experiment KATRIN \cite{katrin}, it is
conceivable that more efforts could be devoted to future
generation of experiments with an improved energy resolution as
good as 0.1 eV.
\begin{table}[h]
\caption{\label{table:1} Relic neutrino capture rate for 100 g of
$^3$H, for a standard Fermi-Dirac distribution with $T_\nu =1.7
\cdot 10^{-4}$ eV (FD). Results are also shown for a Navarro Frenk
and White profile (NFW) and for present day mass distribution of
the Milky Way (MW) for two values of $m_\nu$.}
\begin{center}
\lineup
\begin{tabular}{*{7}{l}} \br $m_\nu$ (eV)  & FD (events yr$^{-1}$)&
NFW (events yr$^{-1}$)& MW (events yr$^{-1}$)\\
\mr 0.3 & 7.5 & 23 & 33 \\ \mr
0.15 & 7.5 & 10 & 12 \\
\br
\end{tabular}
\end{center}
\end{table}
\section{Conclusion}
In this paper we have summarized the analysis of NCB performed in
\cite{cocco}. These processes have the remarkable property of
having no energy threshold on the incoming neutrino energy and
thus they might represent a good and numerous class of
interactions suitable for low energy neutrino flux measurements.
The possibility to pursue the ultimate goal of cosmological relic
neutrino background detection via a future experimental
implementation of this approach depends upon two crucial issues, a
high value for the expected order of magnitude of NCB event rate,
as well a very good energy resolution of the outgoing electron or
positron, of the order of neutrino mass. Both these aspects should
be optimized by a careful choice of the $\beta$ decaying nuclei.
\section*{References}
\end{document}